\documentclass[prl,showpacs,twocolumn,superscriptaddress]{revtex4-1}

\usepackage{amsfonts,amsmath,amssymb,graphicx}
\usepackage[usenames,dvipsnames]{xcolor}
\usepackage{pdfpages}
\usepackage[normalem]{ulem}

\newcommand{\e}{\mathrm{e}}
\newcommand{\aop}{\hat a}
\newcommand{\adop}{\hat a^\dagger}
\newcommand{\aR}{\hat a_\mathrm{out}}
\newcommand{\aL}{\hat a_\mathrm{in}}
\newcommand{\Iop}{\hat I}
\newcommand{\Aop}{\hat b_\mathrm{out}}
\newcommand{\hX}{\hat X}
\newcommand{\hXout}{\hat X_\mathrm{out}}

\newcommand{\hY}{\hat Y}
\newcommand{\hYout}{\hat Y_\mathrm{out}}

\newcommand{\op}[1]{\hat{#1}}
\newcommand{\hH}{\hat{H}}
\newcommand{\hI}{\hat{I}}
\newcommand{\wac}{\omega_\mathrm{ac}}
\newcommand{\wm}{\omega_m}

\newcommand{\vac}{V_\mathrm{ac}}
\newcommand{\vdc}{V_\mathrm{dc}}
\newcommand{\dd}{\mathrm{d}}
\newcommand{\braket}[1]{\langle #1\rangle}
\newcommand{\half}{\frac12}

\begin{document}

\title{Quantum optics theory of electronic noise in coherent conductors}

\author{Farzad Qassemi}
\affiliation{D\'epartment de Physique, Universit\'e de Sherbrooke, 2500 boulevard de l'Universit\'e, Sherbrooke, Qu\'ebec J1K 2R1, Canada}

\author{Arne L. Grimsmo}
\affiliation{D\'epartment de Physique, Universit\'e de Sherbrooke, 2500 boulevard de l'Universit\'e, Sherbrooke, Qu\'ebec J1K 2R1, Canada}

\author{Bertrand Reulet}
\affiliation{D\'epartment de Physique, Universit\'e de Sherbrooke, 2500 boulevard de l'Universit\'e, Sherbrooke, Qu\'ebec J1K 2R1, Canada}

\author{Alexandre Blais}
\affiliation{D\'epartment de Physique, Universit\'e de Sherbrooke, 2500 boulevard de l'Universit\'e, Sherbrooke, Qu\'ebec J1K 2R1, Canada}
\affiliation{Canadian Institute for Advanced Research, Toronto, Canada}

\begin{abstract}
We consider the electromagnetic field generated by a coherent conductor in which electron transport is described quantum mechanically. We obtain an input-output relation linking the quantum current in the conductor to the measured electromagnetic field. This allows us to compute the outcome of measurements on the field in terms of the statistical properties of the current. We moreover show how under ac-bias the conductor acts as a tunable medium for the field, allowing for the generation of single- and two-mode squeezing through fermionic reservoir engineering. These results explain the recently observed squeezing using normal tunnel junctions [G.~Gasse {\it et al.}, Phys.~Rev.~Lett.~{\bf  111} 136601 (2013); J.-C. Forgues {\it et al.}, Phys.~Rev.~Lett.Ê~{\bf 114}Ê 130403Ê (2015)].
\end{abstract}

\pacs{
72.70.+m
42.50.Lc
73.23.-b 
42.50.Dv, 
}

\maketitle

More than sixty years ago, Glauber showed that the electromagnetic radiation produced by a classical electrical current is itself classical~\cite{Glauber51,Glauber63}. The situation can however be different in mesoscopic conductors at low temperature. Indeed, in such conductors  electron transport should no longer be considered classical and current is represented by an operator. Because this operator does not commute with itself when evaluated at different times or frequencies, Glauber's results no longer apply. One may then wonder if a ``quantum current'' may generate a non-classical electromagnetic field. This is the central question addressed in this Letter: how does the quantum properties of current in a coherent conductor imprint on the properties of the electromagnetic field it radiates?

This question was partly addressed in Refs.~\cite{Beenakker00,Beenakker04,Lebedev10} where it was shown, for example, that the statistics of photon emitted by a quantum conductor can deviate from the Poissonian statistics of a coherent state. While photon statistics is most naturally revealed by power detection, measurements on quantum conductors are more typically realized with linear (i.e.~voltage) detectors revealing quadratures of the electromagnetic field radiated by the sample. As a result, Refs.~\cite{Beenakker00,Beenakker04,Lebedev10} only partly answer the question.

More recently it was predicted that, under ac-bias, the electromagnetic field radiated by a coherent conductor can be squeezed~\cite{Bednorz13}. The field is then characterized by fluctuations along one of two quadratures being smaller than the vacuum level. This prediction can be surprising since these quantum states of the electromagnetic field are usually associated with the presence of a nonlinear element, such as a Kerr medium in the optical frequency range~\cite{walls:2008a} or a Josephson junction at microwave frequencies~\cite{yurke:1988a}. Nevertheless, squeezing was experimentally observed using a tunnel junction with linear current-voltage characteristics~\cite{Gasse13, forgues:2015a}. Here squeezing results from quantum shot noise of the junction under ac driving. The predictions of Ref.~\cite{Bednorz13} however only consider correlation functions of the current inside the conductor, not the properties of the emitted field that is squeezed and ultimately measured.

In this Letter, instead of focussing on the current in the coherent conductor, we determine the properties of the field that it radiates. We achieve this, using the langage of quantum optics, by deriving an input-output relation~\cite{Yurke84,yurke:2004a,collett:1984a} directly connecting the radiated electromagnetic field to the current. Given that currents and voltages in electrical circuits are nothing more than another representation of electromagnetic fields, the theoretical methods of quantum optics are particularly well suited.  This relation allows us to compute expectation values of the field corresponding to various types of measurements on mesoscopic samples, including power detection and linear quadrature measurements. We then go a step further and consider the fermionic degrees of freedom of the conductor as a bath for the electromagnetic field of a microwave resonator. Tracing out the conductor's degrees of freedom leads to a Lindblad master equation for the electromagnetic field in a squeezed bath and shows how the electrons in the coherent conductor act as an effective medium for the field. This provides clear insight into the incoherent mechanism responsible for squeezing of the field, as well as a way to compare this mechanism with conventional schemes based on coherent interactions with non-linearities.

Our first step is to model the electromagnetic environment of the sample as a semi-infinite transmission line of characteristic impedance $Z_{0}=\sqrt{L_0/C_0}$, with $L_{0}$ and $C_{0}$ the  inductance and capacitance per unit length respectively. The position-dependent flux $\hat{\phi}_\mathrm{tl}(x,t)$ along the transmission line  is~\cite{Yurke84,yurke:2004a,leppakangas:2013a}
\begin{eqnarray}\label{eq:flux}
\hat{\phi}_{\mathrm{tl}}(x,t) 
& = & 
\alpha\int_{0}^{\infty}\frac{\mathrm{d}\omega}{\sqrt{\omega}}
\left(\aL[\omega]e^{-i\omega(t+ x/v)}\right.\nonumber \\
&  & \left.
+\,\aR[\omega]e^{-i\omega(t - x/v)}+\mathrm{h.c.}\right),
\end{eqnarray}
where $v=1/\sqrt{L_{0}C_{0}}$ is the speed of light in the transmission line and $\alpha=\sqrt{\hbar Z_{0}/2}$~\footnote{We use the Fourier transform convention $f(t)= \int_{-\infty}^{\infty}\mathrm{d}\omega e^{-i\omega t}f[\omega]/2\pi$}. The subscripts `$\mathrm{in}$' and `$\mathrm{out}$' denote components moving towards and away from the sample, respectively. The corresponding annihilation operators  satisfy $[\aL[\omega],\aL^\dag[\omega']]=2\pi\delta(\omega-\omega')$ and similarly for $\aR$. Finally, current at position $x$ in the transmission line is given in terms of the flux by $\hat{I}_{\mathrm{tl}}(x,t)=L^{-1}_{0}\partial_{x}\hat{\phi}_{\mathrm{tl}}(x,t)$, while voltage is $\hat{V}_{\mathrm{tl}}(x,t)=\partial_t\hat{\phi}_{\mathrm{tl}}(x,t)$.

With the sample located at $x=0$, current conservation imposes that
\begin{equation}\label{eq:BC}
\hI_\mathrm{s}(t)=-\hat{I}_{\mathrm{tl}}(x=0,t),
\end{equation}
where $\hI_\mathrm{s}$ is the sample's electron current operator in the presence of the transmission line and of classical voltage bias. This equality links the bosonic operators of the line to the fermionic degrees of freedom of the sample. In the frequency domain, this takes the form
\begin{equation}
\aR[\omega]= \aL[\omega]-i\sqrt{\frac{2Z_{0}}{\hbar\omega}}\hI_\mathrm{s}[\omega] 
\label{eq:BC-general}
\end{equation}
which relates the field travelling away from the conductor $\aR$ to the incoming field $\aL$ and the condutor's current operator $\hI_\mathrm{s}$. This is akin to an input-output boundary condition in quantum optics~\cite{Yurke84,yurke:2004a,Gardiner84}. An expression similar to Eq.~\eqref{eq:BC-general} can be found in Ref.~\cite{Beenakker00} for the case of a quantum conductor coupled to the electromagnetic field freely propagating in three dimensions.

Since $\hI_\mathrm{s}[\omega]$ depends on the current evaluated at all times, it does not commute with $\aL[\omega]$. Care must therefore be taken when evaluating moments of $\aR[\omega]$. In Ref.~\cite{Beenakker00}, this problem was avoided by neglecting the influence of the field's vacuum fluctuations on the current $\hI_\mathrm{s}$. This is justified for a sample of impedance much larger than $Z_0$, thus very poorly matched to the transmission line, and does not correspond to usual experimental conditions where impedance matching is preferable. Here, we address the problem of non-commutativity by writing Eq.~\eqref{eq:BC-general} in terms of the quantum conductor's bare current operator $\hat{I}$ in the absence of the electromagnetic environment, rather than the full current $\hI_\mathrm{s}$ containing the influence of the field. This is done by going to the Heisenberg picture and solving for the current operator perturbatively in the light-matter coupling $\alpha$. This linear response treatment is justified for typical low impedance electromagnetic environments such that $Z_0\ll R_K$ with $R_K=h/e^2 \sim 26~\mathrm{k\Omega}$ the quantum of resistance. For the common experimental value $Z_0=50~\Omega$, one indeed has $e\alpha/\hbar=\sqrt{\pi Z_0/R_K}\sim 0.08 \ll 1$. For low impedance sample and transmission line and when the sample can be treated in the lumped-element limit, we take the interaction between the line's and samples's degrees of freedom to be of the form $H_{\mathrm{I}}(t)=\hI(t)\hat{\phi}_{\mathrm{tl}}(x=0,t)$~\cite{SM}. To first order in $e\alpha/\hbar$ we then find~\cite{graham:1989a}
\begin{equation}\label{eq:BC-linear}
\hI_\mathrm{s}[\omega]=\hI[\omega]+ \hat{V}_{\mathrm{tl}}[\omega]/Z[\omega],
\end{equation}
with $Z[\omega]$ the impedance of the sample. In this expression, $\hI[\omega]$ is the Fourier transform of $\hI(t)$, the electronic current operator evolving according to the bare quantum conductor Hamiltonian~\cite{SM}. In principle, this free Hamiltonian can contain disorder, interactions, etc., as well as the effect of the classical dc and ac bias voltage, $\vdc + \vac \cos{\wac t}$, applied to the conductor.

Combining Eqs.~\eqref{eq:BC-general} and \eqref{eq:BC-linear} directly leads to
\begin{equation}\label{eq:InOut}
\aR[\omega]= r \aL[\omega]-it\frac{\hI[\omega]}{\sqrt{2\hbar\omega Z^{-1}}},
\end{equation}
with $r=\frac{Z-Z_{0}}{Z+Z_{0}}$ the reflection coefficient and $t=\frac{2\sqrt{ZZ_{0}}}{Z+Z_{0}}$ the transmission coefficient with $\left| r \right|^2+\left| t \right|^2=1$. In contrast to Eq.~\eqref{eq:BC-general}, the bare current operator $\hI$ entering Eq.~\eqref{eq:InOut} commutes at all times with the incoming field $\aL$ that has not yet interacted with the conductor. Arbitrary correlation functions of the outgoing field can thus easily be evaluated with this input-output boundary condition. 

As examples, we now discuss the results for different types of common measurements. For simplicity we restrict the discussion to the practically important case of an ideally matched sample, $Z[\omega] = R= Z_{0}$. Then $r=0$ and the outgoing field takes the simple form $\aR[\omega] = -i\hat{I}[\omega]/\sqrt{2 S_\text{vac}(\omega)}$ where $S_\text{vac}(\omega) =  \hbar \omega/R$ is the current noise spectral density of vacuum noise. Measurable properties of the output field are then fully determined by the current. In particular, second order moments of the output field are given in terms of current-current correlation functions which under ac excitation obey~\cite{gabelli:2007a,Gabelli08,SM}
\begin{equation}\label{eq:CurrentCurrentCorrelator}
\begin{split}
 \braket{I[\omega']I[\omega]} 
 &= 2\pi [\tilde{S}(\omega') + S_\text{vac}(\omega')] \delta(\omega'+\omega)\\
& + 2\pi \sum_{p\neq 0} X(\omega') \delta(\omega'+\omega-p\wac).
\end{split}
\end{equation}
In this expression,
\begin{equation}\label{eq:photoassistednoise}
\tilde{S}(\omega) = \sum_{n=-\infty}^\infty J^2_n\left(\frac{e\vac}{\hbar\wac}\right) S\left(\vdc +\frac{n\hbar\wac}{e},\omega\right),
\end{equation}
is the photo-assisted noise, $S(V,\omega) = F\left[S_0(V+\hbar\omega/e)+S_0(V-\hbar\omega/e)\right]/2 +(1-F)S_0(\hbar\omega/e)$ the noise spectral density of current fluctuations in the conductor, $S_0(V) = R^{-1}eV\coth(eV/2k_BT)$ and $F$ the Fano factor~\cite{blanter:2000a}. Moreover,
\begin{equation}\label{eq:Xomega}
\begin{split}
X(\omega) &= \frac{F}{2} \sum_n J_n J_{n+p}\left[S_0\left(\vdc + \frac{\hbar}{e}(\omega+n\wac)\right)\right.\\&
\left.\quad+ (-1)^pS_0\left(\vdc - \frac{\hbar}{e}(\omega+n\wac)\right)\right],
\end{split}
\end{equation}
characterizes the noise dynamics~\cite{gabelli:2008b}. For brevity we have here omitted the argument of the Bessel functions $J_n$ that is the same as in Eq.~\eqref{eq:photoassistednoise}.

We first consider photodetection of the output field in the experimentally relevant situation where the signal is band-pass filtered before detection. This can be taken into account by defining a filtered output field
\begin{equation}\label{eq:A_filtered}
\begin{split}
\Aop(t) 
& = \frac{1}{2\pi\sqrt{B}} \int_B  \dd \omega\, \e^{-i(\omega-\omega_0) t}\aR[\omega],
\end{split}
\end{equation}
where $B$ refers to a measurement bandwidth centered at the observation frequency $\omega_0 \gg 2\pi B$. With this definition, the filtered photo-current is~\cite{SM}
\begin{equation}
\langle \Aop^\dag(t)\Aop(t) \rangle
= \frac{\tilde{S}(\omega_0) - S_\text{vac}(\omega_0)}{2S_\text{vac}(\omega_0)},
\end{equation}
where we have assumed a small filter bandwidth and dropped terms rotating at $\wac$ or faster. As expected, a photodetector is sensitive to the spectral density of the current noise \emph{emitted} by the conductor~\cite{LL97,gavish:2000a,clerk:2010a}. In practice, this can be measured by separating the emission and absorption noise~\cite{aguado:2000a,Deblock03}. A more common detection scheme is to measure the time-averaged power of the emitted electromagnetic field. Again assuming a small measurement bandwidth, we find from Eq.~\eqref{eq:flux} that
$\overline{\braket{V(t)^2}}/R = (2\pi)^2 B \tilde{S}(\omega_0) R$,
where we have omitted a contribution from the vacuum noise of the in-field~\cite{SM}. In contrast to photodetection, measurement of the power of the electromagnetic field is related to the  \emph{symmetric} current-current correlator containing both emission and absorption~\cite{clerk:2010a}. This is not in contradiction with the fact that a passive detector cannot detect vacuum fluctuations~\cite{gavish:2000a}. Power measurements are indeed performed using active devices like amplifiers and mixers.

Following the experiments of Refs.~\cite{Gasse13,forgues:2015a}, we now consider measurement of field quadratures as obtained by homodyne detection~\cite{walls:2008a}. Defining quadratures of the output field in the frequency domain as $\hXout[\omega] = \aR^\dagger[\omega] + \aR[\omega]$ and $\hYout[\omega] = i(\aR^\dag[\omega] - \aR[\omega])$, and using Eq.~\eqref{eq:InOut}, we immediately find for the variance of these quantities~\cite{SM}
\begin{equation} \label{eq:QuadraturesOut}
\begin{split}
  \Delta \hXout^2[\omega] &= \frac{\braket{\{\op{I}[-\omega],\op{I}[\omega]\}} - 2\braket{\op{I}[\omega]^2}}{2S_\text{vac}(\omega)},\\
  \Delta \hYout^2[\omega] &= \frac{\braket{\{\op{I}[-\omega],\op{I}[\omega]\}} + 2\braket{\op{I}[\omega]^2}}{2S_\text{vac}(\omega)}.
\end{split}
\end{equation}
In practice, $\braket{I[\omega]^2}$ is only non-zero in the presence of ac-bias on the sample. Indeed, as expressed by Eq.~\eqref{eq:CurrentCurrentCorrelator}, modulation of the bias voltage at frequency $\wac$ induces correlations between Fourier components of the current separated by $p\wac$, with $p$ an integer~\cite{gabelli:2007a,Gabelli08}. For $p\wac = 2\omega_0$, and defining filtered output quadratures, $\hat{X}_\mathrm{out,f}(t) = \Aop^\dagger(t) + \Aop(t)$ and $\hat{Y}_\mathrm{out,f}(t) = i[\Aop^\dagger(t) - \Aop(t)]$, we find~\cite{SM}
\begin{equation} \label{eq:XY_filtered}
\begin{split}
   \Delta \hat{X}_\mathrm{out,f}^2(t) &= 2 \left(N(\omega_0) + \half - M(\omega_0)\right),\\
  \Delta \hat{Y}_\mathrm{out,f}^2(t)  &= 2 \left(N(\omega_0) + \half + M(\omega_0)\right),
\end{split}
\end{equation}
where
\begin{equation}
  N(\omega) = \frac{\tilde{S}(\omega)-S_\text{vac}(\omega)}{2 S_\text{vac}(\omega)},\quad
  M(\omega) = \frac{X(\omega)}{2S_\text{vac}(\omega)}.\label{eq:NMdef}
\end{equation}
Clearly, the $X$ quadrature of the output field is squeezed when $M(\omega_0)> N(\omega_0)$, equivalently $X(\omega_0)>\tilde{S}(\omega_0)-S_\text{vac}(\omega_0)$, with $M(\omega)$ and $N(\omega)$ bounded from the Heisenberg inequality by $N(\omega)[N(\omega)+1]\ge M(\omega)^2$~\cite{walls:2008a}. The same condition for squeezing was found in Refs.~\cite{Gasse13,Bednorz13} by directly postulating the link between the field quadratures and the current operator for a normal tunnel junction.  The squeezing generated by such a junction is however moderate. At zero temperature we expect maximum squeezing of $\sim2$ dB while the experiment of Ref.~\cite{Gasse13} reported squeezing of 1.3 dB.

\begin{figure}
\includegraphics[width=0.95\columnwidth]{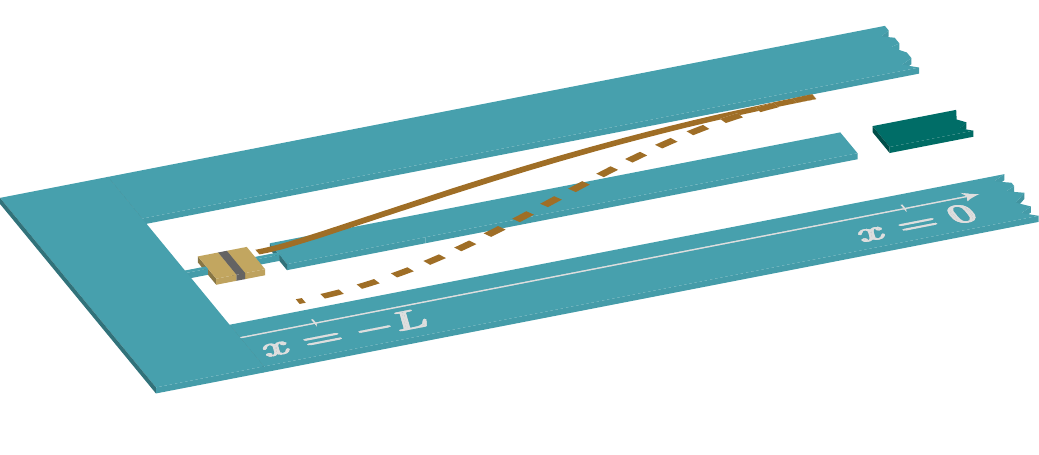}
\caption{Transmission line resonator (blue) terminated by a normal tunnel junction. The first resonator mode envelope are illustrated for $R/Z_0=0.1$ (full brown line) and $R/Z_0=2000$ (dashed brown line). The output field can be measured via the capacitive coupling to the output port (dark blue).}
\label{figCavity}
\end{figure}

To better understand the mechanism responsible for squeezing by the quantum conductor we now derive an equation of motion for the state $\rho(t)$ of the field. This is done by following the standard quantum optics approach: the bath is integrated out invoking the Born-Markov approximation to obtain a Lindblad master equation describing the dynamics of the field only~\cite{walls:2008a}. The crucial difference from the usual treatment is that here the fermionic degrees of freedom of the sample play the role of bath for the bosonic modes of the field.  Moreover, it is possible to engineer the system-bath interaction with the ac modulation frequency, leading to different field steady-states.

To simplify the discussion and because it is experimentally relevant~\cite{souquet:2014a,altimiras:2014a,parlavecchio:2015a}, we consider the setup illustrated in Fig.~\ref{figCavity} where a normal tunnel junction is fabricated at the end of a $\lambda/4$ transmission line-resonator of characteristic impedance $Z_0$. The case of a Josephson junction has been considered in Refs.~\cite{hofheinz:2011a,leppakangas:2013a,gramich:2013a,armour:2013a}. The effect of the junction on the resonator is easily found by decomposing the resonator flux in terms of normal modes $\phi(x,t) = \sum_m\phi_m(t) u_m(x)$, with $u_m(x)$ the mode envelope~\cite{SM,bourassa:2012a}. The full line in Fig.~\ref{figCavity} illustrates $|u_1(x)|$ for a junction impedance $R<Z_0$, while the dashed line corresponds to $u_1(x)$ for $Z_0>R$. As expected, in the former case the junction acts as a short to ground and the mode envelope approaches that of a $\lambda/4$ resonator, except for a small gap $|u_1(-L)|\sim R/Z_0$ at the location of the junction. On the other hand, for large tunnel resistance the junction acts as an open and the resonator's bias on the junction is larger with $|u_1(-L)|\sim1+(Z_0/R)^2$.

Having characterized the resonator mode in the presence of the junction, we now obtain the master equation assuming $Z_0\ll R_K$. In this limit, the interaction  Hamiltonian reads $\hH_\mathrm{I} = \sum_m \alpha_m \Iop (\adop_m+\aop_m)$, where $\alpha_m = \sqrt{\hbar Z_m}u_m(-L)$ with $Z_m$ the effective impedance of the resonator's $m\mathrm{th}$ mode and $\hat a^{(\dag)}_m$ the annihilation (creation) operator for the same mode~\cite{SM}. As above, the current operator $\Iop$ takes into account the presence of classical dc and ac bias on the junction.

We first focus on the situation where the ac frequency is, as above, $p\wac=2\wm$ where $p$ is an integer and $\wm$ now the frequency of the $m$th resonator mode. In the rotating-wave approximation we find~\cite{SM}
\begin{equation}\label{eq:master:singlemodesqueezing}
  \begin{split}
    \dot{\rho}(t) 
    & = \kappa_m (N_m+1) \mathcal{D}[\aop_m]\rho + \kappa_m N_m\mathcal{D}[\adop_m]\rho 
    \\ &+ \kappa_m M_m\mathcal{S}[\aop_m]\rho + \kappa_m M_m\mathcal{S}[\adop_m]\rho,
  \end{split}
\end{equation}
with $\mathcal{D}[\aop]\rho = \aop\rho \adop - \{\adop\aop,\rho\}/2$ and $\mathcal{S}[X]\rho = \aop \rho \aop - \{\aop^2,\rho\}/2$. Eq.~\eqref{eq:master:singlemodesqueezing} is the standard master equation of a bosonic mode in a squeezed bath~\cite{walls:2008a} where $\kappa_m = u_m(-L)^2\omega_m Z_m/R$ is the cavity damping rate caused by the tunnel junction resistance~\footnote{The Markov approximation is valid when the environment's time scale is short with respect to $\kappa^{-1}$~\cite{Carmichael:1999aa} or in other words for $\kappa^{-1}\gg \mathrm{Min}[\hbar/k_BT,\hbar/e\vdc]$. As expected, this implies that the sample resistance should not be matched to $Z_0$.}. The thermal photon number $N_m = N(\omega_m)$ and the quantity $M_m = M(\omega_m)$  responsible for squeezing are the same as in Eq.~\eqref{eq:NMdef}. Evolution under Eq.~\eqref{eq:master:singlemodesqueezing} leads to steady-state variances of the intracavity quadratures $\hX_m=\adop_m+\aop_m$ and $\hY_m=i(\adop_m-\aop_m)$ taking the form $\Delta X_m^2 = 2N_m + 1 - 2M_m$ and $\Delta Y_m^2 = 2N_m + 1 + 2M_m$. In other words, intra-cavity squeezing is identical to what was found in Eq.~\eqref{eq:XY_filtered} in the absence of the resonator.

The form of the above master equation clearly illustrates the dissipative nature of squeezing by a tunnel junction. This type of squeezing by dissipation has been explored theoretically in various systems and, in particular, in Ref.~\cite{Didier14} where it was shown that modulating the quality factor of a linear cavity could lead to ideal and unbounded squeezing. A similar mechanism is in action here with the periodic modulation of the Fermi level of the tunnel junction by the ac bias. The achievable squeezing is however neither pure nor unbounded, with the purity $p=S_\mathrm{vac}/\sqrt{\tilde S^2-X^2}<1$. At zero temperature, the highest expected purity is $p\sim0.91$ corresponding to the 2 dB of squeezing mentioned above. This conclusion also applies to the cavity output field. Indeed, taking into account an output port (illustrated by the capacitor on the right-hand-side of Fig.~\ref{figCavity}) reduces intracavity squeezing by adding vacuum noise. This additional contribution is however absent from the cavity output field if the decay rates at the two ends of the resonator are matched~\cite{collett:1984a}, leaving the degree of squeezing unchanged from the above input-output theory without cavity~\cite{SM}.

Taking advantage of the multi-mode structure of the resonator, other choices of ac drive can lead to entangled steady-states. In particular, taking $p\wac=\wm+\omega_n$ results in~\cite{SM}
\begin{equation}\label{eq:master:twomodesqueezing}
  \begin{split}
    \dot{\rho}(t) = &\sum_{l=n,m} \left\{\kappa_l (N_l+1) \mathcal{D}[\aop_l]\rho + \kappa_l N_l\mathcal{D}[\adop_l]\rho \right\} \\
     &+\sqrt{\kappa_n\kappa_m}M_{nm} \left(\aop_n\rho\aop_m+\aop_m\rho\aop_n + \{\aop_n\aop_m,\rho\}\right)\\
     &+\sqrt{\kappa_n\kappa_m}M_{nm} \left(\adop_n\rho\adop_m+\adop_m\rho\adop_n + \{\adop_n\adop_m,\rho\}\right),
  \end{split}
\end{equation}
where $\kappa_l$ and $N_l$ are the same as above and $M_{nm} = X(\omega_n)/2\sqrt{S_\text{vac}(\omega_n)S_\text{vac}(\omega_m)}$. This master equation leads to two-mode squeezing. Indeed, in steady-state the variance of the joint quadratures $\hX_\pm = \hX_n \pm \hX_m$ and $\hY_\pm = \hY_n \pm \hY_m$ are
\begin{equation}
  \begin{split}
  \Delta X_{+}^2 = \Delta Y_{-}^2 = 2( N_n +  N_m + 1 - 2M_{nm}),\\
  \Delta X_{-}^2 = \Delta Y_{+}^2 = 2( N_n +  N_m + 1 + 2M_{nm}),
  \end{split}
\end{equation}
where we have assumed $\kappa_n = \kappa_m$ for simplicity. Similarly to the above single-mode case, the pairs of commuting quadratures $\Delta X_{+}^2$ and $\Delta Y_{-}^2$ are squeezed for $2M_{nm} > N_n +  N_m$. It is interesting to point out that these quadratures are entangled when $\Delta X_+^2 + \Delta Y_-^2 < 4$~\cite{duan:2000a}. This type of two-mode squeezing generated by a normal tunnel junction under the above ac modulation frequency was already experimentally reported in Ref.~\cite{forgues:2015a}.

We finally consider the situation where the ac modulation is such that  $p\wac = |\omega_{n}-\omega_{m}|$ in which case the master equation takes the form~\cite{SM}
\begin{equation}\label{eq:master:differencefrequency}
  \begin{split}
    \dot{\rho}(t) = &\sum_{l=n,m} \left\{\kappa_l (N_l+1) \mathcal{D}[\aop_l]\rho + \kappa_l N_l\mathcal{D}[\adop_l]\rho \right\} \\
     &+\sqrt{\kappa_n\kappa_m}M_{nm} \left(\aop_n\rho\adop_m+\adop_m\rho\aop_n - \{\adop_m\aop_n,\rho\}\right)\\
     &+\sqrt{\kappa_n\kappa_m}M_{nm} \left(\adop_n\rho\aop_m+\aop_m\rho\adop_n - \{\adop_n\aop_m,\rho\}\right).
  \end{split}
\end{equation}
Rather than two-mode squeezing, this describes correlated decay where emission by mode $n$ stimulates emission from mode $m$, and vice-versa. Under this evolution, the variance of the above joint quadratures keep the same form, except for $\Delta X_{+}^2$ and $\Delta X_{-}^2$ whose role are exchanged. Since $[\hX_-,\hY_-] = 4i$, the variance of these two quadratures must respect $\Delta X_- \Delta Y_- \ge 2$ implying that $N_n+N_m\ge 2M_{nm}$. In other words, these quadratures cannot be squeezed below the vacuum level, also implying that the two modes are not entangled, and the master equation Eq.~\eqref{eq:master:differencefrequency} only leads to squashing.

In summary, we have derived an input-output relation linking properties of the electrons in a quantum conductor to the measured electromagnetic field emitted by the conductor. We have also shown how the conductor act as a tunable medium for the field, allowing for the generation of single- and two-mode squeezing through fermionic reservoir engineering. Recent experimental observations of squeezing produced by a tunnel junction can be understood within this framework.

{\it Note added.} Recently, we became aware of an alternate description of squeezing by tunnel junction in a resonator~\cite{mendes:2015a}.

{\it Acknowledgements-- }
We thank Julien Gabelli and Karl Thibault for useful discussions.
This work was supported by the Canada Excellence Research Chairs program, NSERC, FRQNT via INTRIQ and the Universit\'e de Sherbrooke via EPIQ.


%

\clearpage


\section*{Supplementary material}

# Supplemental Material for "Quantum optics theory of electronic noise in coherent conductors"

Farzad Qassemi,[1] Arne L. Grimsmo,[1] Bertrand Reulet,[1] and Alexandre Blais[1, 2]

[1]*Département de Physique, Université de Sherbrooke,*
*2500 boulevard de l'Université, Sherbrooke, Québec J1K 2R1, Canada*
[2]*Canadian Institute for Advanced Research, Toronto, Canada*

## I. NOISE PROPERTIES OF AN AC-BIASED TUNNEL JUNCTION

For concreteness we consider here a normal tunnel junction. The results of this section are however general and have been obtained for an arbitrary quantum conductor using the Landauer-Büttiker formalism in Refs. [1, 2]. Our starting point is therefore the Hamiltonian of a tunnel junction biased by a classical voltage $V_{\rm dc} + V_{\rm ac} \cos \omega_{\rm ac} t$ and by the transmission line voltage at the position $x = 0$ of the junction, described by the operator $\hat{V}_{\rm tl}(t)$,

$$\hat{H} = \hat{H}_{\rm tj} + \hat{H}_{\rm T}, \tag{1}$$

where

$$\hat{H}_{\rm tj} = \hat{H}_{\rm tj,0} + \hat{H}_{\rm tj,1}, \tag{2}$$

$$\hat{H}_{\rm tj,0} = \sum_k \left( \epsilon_k + eV_{\rm dc} \right) \hat{c}_k^\dagger \hat{c}_k + \sum_q \epsilon_q c_q^\dagger c_q, \tag{3}$$

$$\hat{H}_{\rm tj,1} = \sum_k \left[ eV_{\rm ac} \cos(\omega_{\rm ac} t) + e\hat{V}_{\rm tl}(t) \right] \hat{c}_k^\dagger \hat{c}_k, \tag{4}$$

$$\hat{H}_T = \sum_{kq} t_{kq} \hat{c}_k^\dagger \hat{c}_q + {\rm H.c.}. \tag{5}$$

$\hat{H}_{\rm tj,0}$ is the junction Hamiltonian in the presence of the dc voltage, $\hat{H}_{\rm tj,1}$ is the contribution coming from the ac voltage bias as well as the transmission line voltage, and $\hat{H}_T$ is the tunneling Hamiltonian. The operators $\hat{c}_k$ and $\hat{c}_q$ are fermionic annihilation operators at the two different leads of the junction.

The voltage operator, $\hat{V}_{\rm tl}(t) = {\rm d}\hat{\phi}_{\rm tl}(t)/{\rm d}t$ for the transmission line is the time-derivative of the transmission line flux, evaluated at the position of the junction, with

$$\hat{\phi}_{\rm tl}(x,t) = \alpha \int_0^\infty \frac{{\rm d}\omega}{\sqrt{\omega}} \left( \hat{a}_{\rm in}[\omega] e^{-i\omega(t + x/v)} + \hat{a}_{\rm out}[\omega] e^{-i\omega(t - x/v)} + {\rm H.c.} \right), \tag{6}$$

where $\alpha = \sqrt{\hbar Z_0 / 2}$. Following the standard step, a unitary transformation is applied leading to $\hat{H} \to \hat{H}_I = \hat{U}\hat{H}\hat{U}^\dagger - i\hat{U}\frac{{\rm d}}{{\rm d}t}\hat{U}^\dagger$, with

$$\hat{U}(t) = \exp\left[ i\left( \frac{t}{\hbar}\hat{H}_{\rm tj,0} + \frac{eV_{\rm ac}}{\hbar\omega_{\rm ac}} \sin(\omega_{\rm ac} t)\hat{c}_k^\dagger \hat{c}_k + \frac{e}{\hbar}\hat{\phi}_{\rm tl}(t) \right) \right]. \tag{7}$$

This removes $H_{\rm tj}$ and leads to

$$
\begin{aligned}
\hat{H}_I(t) &= \sum_{kq} \exp\left[ \frac{i}{\hbar}\left( eV_{\rm dc} + \epsilon_k - \epsilon_q \right) t + \frac{ie}{\hbar}\left( \frac{V_{\rm ac}}{\omega_{\rm ac}} \sin(\omega_{\rm ac} t) + \hat{\phi}_{\rm tl}(t) \right) \right] \hat{c}_k^\dagger \hat{c}_q + {\rm H.c.} \\
&= \sum_{n=-\infty}^{\infty} \sum_{kq} J_n(A) t_{kq} \exp\left[ \frac{i}{\hbar}\left( eV_{\rm dc} + \epsilon_k - \epsilon_q + \hbar n\omega_{\rm ac} \right) t + \frac{ie}{\hbar}\hat{\phi}_{\rm tl}(t) \right] \hat{c}_k^\dagger \hat{c}_q + {\rm H.c.},
\end{aligned}
\tag{8}
$$

where $J_n(A)$ are Bessel functions of the first kind and $A = eV_{\rm ac}/\hbar\omega_{\rm ac}$. Taking advantage of the fact that

$$\sqrt{\frac{e^2 Z_0}{2\hbar}} = \sqrt{\frac{2\pi Z_0}{2R_K}} \ll 1, \tag{9}$$

where $R_{\mathrm{K}} = h/e^2 \sim 25$ k$\Omega$ is the quantum of resistance, we expand the exponential in (8) to first order in $e\hat{\phi}_{\mathrm{tl}}(t)/\hbar$

$$\hat{H}_I(t) \approx \sum_{n=-\infty}^{\infty} \sum_{kq} J_n(A) t_{kq} \mathrm{e}^{\frac{i}{\hbar}(eV_{\mathrm{dc}}+\epsilon_k-\epsilon_q+\hbar n\omega_{\mathrm{ac}})t} \left(1 + \frac{ie}{\hbar}\hat{\phi}_{\mathrm{tl}}(t)\right) \hat{c}_k^{\dagger}\hat{c}_q + \mathrm{H.c..} \tag{10}$$

The first term in the parenthesis of the above equation can easily be transformed away with a further unitary transformation, leaving us with

$$\hat{H}_I(t) \approx \hat{I}(t)\hat{\phi}_{\mathrm{tl}}(t), \tag{11}$$

where we have introduced the standard current operator of the junction in the presence of ac bias

$$\hat{I}(t) = \frac{ie}{\hbar} \sum_{n=-\infty}^{\infty} \sum_{kq} J_n(A) t_{kq} \mathrm{e}^{\frac{i}{\hbar}(eV_{\mathrm{dc}}+\epsilon_k-\epsilon_q+\hbar n\omega_{\mathrm{ac}})t} \hat{c}_k^{\dagger}\hat{c}_q + \mathrm{H.c..} \tag{12}$$

As discussed in the main paper, second order moments of the output field are determined by the current-current correlation function $\langle \hat{I}[\omega']\hat{I}[\omega]\rangle$, where $\hat{I}[\omega] = \int \mathrm{d}t \exp(i\omega t)\hat{I}(t)$ is the Fourier transformed junction current. From (12) we find

$$\langle \hat{I}[\omega']\hat{I}[\omega]\rangle = 2\pi \sum_{p=-\infty}^{\infty} X_+^{(p)}(\omega')\delta(\omega'+\omega-p\omega_{\mathrm{ac}}), \tag{13}$$

where

$$X_+^{(p)}(\omega) = \frac{1}{2} \sum_n J_n\left(\frac{eV_{\mathrm{ac}}}{\hbar\omega_{\mathrm{ac}}}\right) J_{n+p}\left(\frac{eV_{\mathrm{ac}}}{\hbar\omega_{\mathrm{ac}}}\right) \left[S^{\mathrm{LR}}(\omega+n\omega_{\mathrm{ac}}) + (-1)^p S^{\mathrm{RL}}(\omega+n\omega_{\mathrm{ac}})\right], \tag{14}$$

with

$$\begin{aligned} S^{\mathrm{LR}}(\omega) &= 2\frac{e^2}{\hbar} \sum_{kq} |t_{kq}|^2 2\pi\delta(eV_{\mathrm{dc}}+\epsilon_k-\epsilon_q+\hbar\omega)f_k|1-f_q| \\ &= S_0(\hbar\omega/e+V_{\mathrm{dc}}) + S_{\mathrm{vac}}(\omega+eV_{\mathrm{dc}}/\hbar) \end{aligned} \tag{15}$$

and

$$\begin{aligned} S^{\mathrm{RL}}(\omega) &= 2\frac{e^2}{\hbar} \sum_{kq} |t_{kq}|^2 2\pi\delta(eV_{\mathrm{dc}}+\epsilon_k-\epsilon_q-\hbar\omega)f_q|1-f_k| \\ &= S_0(\hbar\omega/e-V_{\mathrm{dc}}) + S_{\mathrm{vac}}(\omega-eV_{\mathrm{dc}}/\hbar). \end{aligned} \tag{16}$$

In these expressions, $f$ is the Fermi-Dirac distribution and we have assumed $t_{kq}$ to be energy independent. Finally, in the last step we have defined

$$S_0(V) = \frac{eV}{R} \coth\left(\frac{eV}{2k_B T}\right), \tag{17}$$

$$S_{\mathrm{vac}}(\omega) = \frac{\hbar\omega}{R}. \tag{18}$$

with $R^{-1} = 2\pi(e^2/\hbar)|t|^2 d_L d_R$ the tunnel resistance and $d_L$ and $d_R$ the density of modes on the left and the right of the tunnel junction, respectively. Note that $X_+^{(p)}(\omega)$ is a real function for energy-independent transmission. It is straightforward to show that $X_+^{(p)}$ can also be written as

$$X_+^{(0)}(\omega) = \tilde{S}(\omega) + S_{\mathrm{vac}}(\omega), \tag{19}$$

$$X_+^{(p\neq 0)}(\omega) = X(\omega), \tag{20}$$

with

$$\tilde{S}(\omega) = \frac{1}{2} \sum_n J_n(A)^2 \left[S_0\left(V_0+\frac{\hbar}{e}(\omega+n\omega_{\mathrm{ac}})\right) + S_0\left(V_0-\frac{\hbar}{e}(\omega+n\omega_{\mathrm{ac}})\right)\right], \tag{21}$$

$$X(\omega) = \frac{1}{2} \sum_n J_n\left(\frac{eV_{\mathrm{ac}}}{\hbar\omega_{\mathrm{ac}}}\right) J_{n+p}\left(\frac{eV_{\mathrm{ac}}}{\hbar\omega_{\mathrm{ac}}}\right) \left[S_0\left(V_{\mathrm{dc}}+\frac{\hbar}{e}(\omega+n\omega_{\mathrm{ac}})\right) + (-1)^p S_0\left(V_{\mathrm{dc}}-\frac{\hbar}{e}(\omega+n\omega_{\mathrm{ac}})\right)\right]. \tag{22}$$

As already mentioned, while we have obtained these expressions taking a tunnel junction (Fano factor $F = 1$) as a specific example, they can be obtained for a general quantum conductor using the Landauer-Büttiker formalism [1, 2].

## II. MEASUREMENTS OF THE TRANSMISSION LINE OUTPUT FIELD

### A. Power measurement

From Eq. (1) of the main paper, the voltage at $x = 0$ is

$$\hat{V}(t) = -i\alpha \int_0^\infty \mathrm{d}\omega \sqrt{\omega} \left( \hat{a}_{\mathrm{in}}[\omega] \mathrm{e}^{-i\omega t} + \hat{a}_{\mathrm{out}}[\omega] \mathrm{e}^{-i\omega t} - \mathrm{H.c.} \right). \tag{23}$$

In an experiment, the voltage will be filtered by a bandpass filter before measurement and for this reason we define the filtered voltage

$$\hat{V}_f(t) = -i\alpha \int_B \mathrm{d}\omega \sqrt{\omega} \left( \hat{a}_{\mathrm{in}}[\omega] \mathrm{e}^{-i\omega t} + \hat{a}_{\mathrm{out}}[\omega] \mathrm{e}^{-i\omega t} - \mathrm{H.c.} \right), \tag{24}$$

where the subscript $B$ refers to the measurement bandwidth centred around the observation frequency $\omega_0 \gg 2\pi B$. Squaring and taking the expectation value yields

$$\langle \hat{V}_f(t)^2 \rangle = \alpha^2 \iint_B \mathrm{d}\omega \mathrm{d}\omega' \sqrt{\omega' \omega} \left( \langle \hat{a}_{\mathrm{out}}[\omega'] \hat{a}_{\mathrm{out}}^\dagger[\omega] \rangle \mathrm{e}^{i(\omega - \omega')t} + \langle \hat{a}_{\mathrm{out}}^\dagger[\omega'] \hat{a}_{\mathrm{out}}[\omega] \rangle \mathrm{e}^{i(\omega' - \omega)t} \right.$$
$$\left. - \langle \hat{a}_{\mathrm{out}}[\omega'] \hat{a}_{\mathrm{out}}[\omega] \rangle \mathrm{e}^{-i(\omega + \omega')t} - \langle \hat{a}_{\mathrm{out}}^\dagger[\omega'] \hat{a}_{\mathrm{out}}^\dagger[\omega] \rangle \mathrm{e}^{i(\omega + \omega')t} \right) + 2\pi \alpha^2 \int_B \mathrm{d}\omega \omega \langle \hat{a}_{\mathrm{in}}[\omega] \hat{a}_{\mathrm{in}}^\dagger[\omega] \rangle, \tag{25}$$

where we have assumed an impedance matched junction for which $\hat{a}_{\mathrm{out}}[\omega] = -i\hat{I}[\omega]/\sqrt{2S_{\mathrm{vac}}(\omega)}$. We have also used $[\hat{a}_{\mathrm{in}}, \hat{I}] = 0$, $\langle \hat{a}_{\mathrm{in}}[\omega] \rangle = 0$ (vacuum input). The last term represents the input field vacuum noise contribution and is dropped in the Letter.

The time-averaged power is obtained using Eq. (13) and dropping fast-rotating terms to find

$$\overline{\langle \hat{V}_f(t)^2 \rangle} = \frac{(2\pi)^2 \alpha^2 \omega_0 \tilde{S}(\omega_0)}{B^{-1} S_{\mathrm{vac}}(\omega_0)} + 2\pi \alpha^2 \int_B \mathrm{d}\omega \omega \langle \hat{a}_{\mathrm{in}}[\omega] \hat{a}_{\mathrm{in}}^\dagger[\omega] \rangle, \tag{26}$$

where the overline denotes time-averaging.

### B. Quadratures

We define the filtered output field [3]

$$\hat{b}_{\mathrm{out}}(t) = \frac{1}{2\pi \sqrt{B}} \int_B \mathrm{e}^{-i(\omega - \omega_0)t} \hat{a}_{\mathrm{out}}[\omega], \tag{27}$$

and quadratures

$$\hat{X}_{\mathrm{out,f}}(t) = \hat{b}_{\mathrm{out}}(t) + \hat{b}_{\mathrm{out}}^\dagger(t), \tag{28}$$

$$\hat{Y}_{\mathrm{out,f}}(t) = -i \left( \hat{b}_{\mathrm{out}}(t) - \hat{b}_{\mathrm{out}}^\dagger(t) \right). \tag{29}$$

Repeating a similar calculation as above, we find for the variances

$$\Delta \hat{X}_{\mathrm{out,f}}^2 = 2 \left[ N(\omega_0) + \frac{1}{2} - M(\omega_0) \right], \tag{30}$$

$$\Delta \hat{Y}_{\mathrm{out,f}}^2 = 2 \left[ N(\omega_0) + \frac{1}{2} + M(\omega_0) \right], \tag{31}$$

where we have dropped all fast rotating terms. The $X$-quadrature is squeezed for $M(\omega_0) > N(\omega_0)$, while the $Y$-quadrature is anti-squeezed.

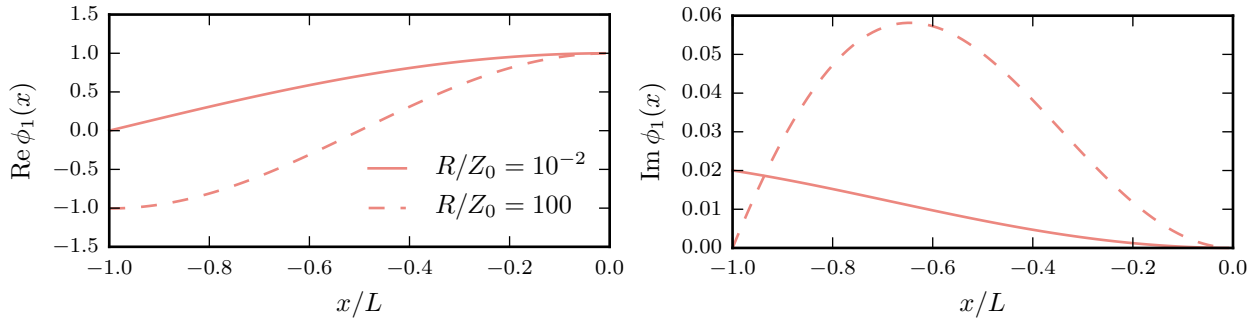

FIG. 1. Real and imaginary part of the lowest frequency mode envelope function, $u_1(x)$, for $R/Z_0 = 10^{-2}$ (solid) and $R/Z_0 = 100$ (dashed).

## III.  TRANSMISSION LINE RESONATOR TERMINATED BY A TUNNEL JUNCTION

In this section we compute the mode functions of a transmission line resonator terminated by a tunnel junction to ground at one end ($x = -L$) and with open boundary conditions at the other end ($x = 0$). This system is illustrated in Fig. 1 of the Letter. We model the junction as a resistor with impedance $R$ assuming for simpicity that the junction capacitance can be neglected.

The resonator has length $L$, capacitance per unit length $c$ and inductance per unit length $l$. Following Ref. [4], we expand the flux at position $x$ of the resonator over modes $m$ as $\phi(x,t) = \sum_m \psi_m(t) u_m(x)$ with $u_m(t)$ oscillating at the mode frequency $\tilde{\omega}_m = \nu \tilde{k}_m$ with $\tilde{k}_m$ the mode wavevector and $\nu = 1/\sqrt{lc}$ the speed of light in the resonator.

The mode envelope $u_m(x)$ is found from the ansatz

$$u_m(x) = \cos(\tilde{k}_m x + \theta_m). \tag{32}$$

The phase $\theta_m$ is fixed by setting the current to be zero at $x = 0$

$$\frac{1}{l} \left. \frac{\partial \phi(x,t)}{\partial x} \right|_{x=0} = 0, \tag{33}$$

This immediately leads to $\theta_m = 0$. On the other hand, at the junction location we have that

$$\frac{1}{l} \left. \frac{\partial \phi(x,t)}{\partial x} \right|_{x=-L} = \frac{1}{R} \frac{\partial \phi(-L,t)}{\partial t} \tag{34}$$

which leads to

$$\frac{-i\tilde{\omega}_m}{R} \cos(\tilde{k}_m L) = \frac{\tilde{k}_m}{l} \sin(\tilde{k}_m L) \quad \Rightarrow \quad \cot(\tilde{k}_m L) = i \frac{R}{Z_0}, \tag{35}$$

where $Z_0 = \sqrt{l/c}$ is the characteristic impedance of the transmission line. This transcendental equation can be solved numerically to find the values of $\tilde{k}_m$. Since the impedance of the junction is real, the wavevectors are in general complex, and the envelope functions are thus complex functions of $x$. Examples of mode envelope functions for respectively large and small tunnel junction resistance $R$ are shown in Fig. 1.

Approximate expressions for the value of the envelope functions at $x = -L$ can be found in the limits of large impedance mismatch between the junction and the transmission line. Focusing on the fundamental mode, *i.e.* the smallest non-zero $k_m$, we find for $Z_0/R \to 0$

$$\tilde{\omega}_1 = \omega_1 - i\kappa_1 \simeq \frac{\pi v}{L} - i\frac{Z_0}{R}\frac{v}{L}, \tag{36}$$

$$u_1(-L) \simeq -\left[1 + \left(\frac{Z_0}{R}\right)^2\right], \tag{37}$$

and for $Z_0/R \to \infty$

$$\tilde{\omega}_1 = \omega_1 - i\kappa_1 \simeq \frac{\pi v}{2L} - i\frac{R}{Z_0}\frac{v}{L}, \tag{38}$$

$$u_1(-L) \simeq i\frac{R}{Z_0}. \tag{39}$$

In the former limit, the boundary condition approaches that of a $\lambda/2$ resonator, *i.e.* open boundary conditions at both ends, and in the latter case that of a $\lambda/4$ resonator, *i.e.* a resonator terminated to ground at $x = -L$ and open at $x = 0$. It is important to note that the coupling between the tunnel junction and the resonator modes can be non-zero also for large $Z_0/R$ due to the non-zero imaginary part of the envelope function. The predictions for the decay rates $\kappa_m$ made with this simplified model is verified by a master equation treatment in the next section.

## IV. MASTER EQUATION: JUNCTION IN A RESONATOR

We now derive a master equation for the resonator modes, assuming weak coupling between the tunnel junction and the transmission line resonator, and invoking the usual Born-Markov approximations. Our starting point is the same as in Sec. I, only that the semi-infinite transmission line is now replaced by a transmission-line resonator. We denote the resonator's voltage operator by $\hat{V}(t)$ and write the total Hamiltonian as

$$\hat{H} = \sum_m \omega_m \hat{a}_m^\dagger \hat{a}_m + \hat{H}_{\text{tj}} + \hat{H}_{\text{T}}, \tag{40}$$

where

$$\hat{H}_{\text{tj}} = \hat{H}_{\text{tj},0} + \hat{H}_{\text{tj},1}, \tag{41}$$

$$\hat{H}_{\text{tj},0} = \sum_k \left(\epsilon_k + eV_{\text{dc}}\right) \hat{c}_k^\dagger \hat{c}_k + \sum_q \epsilon_q c_q^\dagger c_q, \tag{42}$$

$$\hat{H}_{\text{tj},1} = \sum_k \left[eV_{\text{ac}}\cos(\omega_{\text{ac}}t) + e\hat{V}(t)\right] \hat{c}_k^\dagger \hat{c}_k, \tag{43}$$

$$\hat{H}_T = \sum_{kq} t_{kq} \hat{c}_k^\dagger \hat{c}_q + \text{H.c.}. \tag{44}$$

$\hat{H}_{\text{tj},0}$ is the junction Hamiltonian in the presence of the dc voltage, $\hat{H}_{\text{tj},1}$ is the contribution coming from the ac voltage bias as well as the resonator voltage, while $\hat{H}_T$ is the junction's tunnel Hamiltonian.

Following Sec. I, the voltage operator $\hat{V}(t) = \mathrm{d}\hat{\phi}(t)/\mathrm{d}t$ is related to the transmission line flux evaluated at the position of the junction $x = -L$ with

$$\hat{\phi}(t) \equiv \hat{\phi}(-L, t) = \sum_m \sqrt{\hbar Z_m} u_m(-L) \left(\hat{a}_m \mathrm{e}^{-i\omega_m t} + \hat{a}_m^\dagger \mathrm{e}^{i\omega_m t}\right), \tag{45}$$

where, for simplicity, we have taken the mode envelope functions to be real at the position of the junction and have introduced the effective mode impedance $Z_m = 1/\omega_m L c$ [5].

Taking advantage of the fact that

$$\sqrt{\frac{e^2}{\hbar Z_m}} u_m(-L) = \sqrt{\frac{2\pi Z_m}{R_K}} u_m(-L) \ll 1, \tag{46}$$

and following the same steps as in Sec. I, we again find

$$\hat{H}_I(t) \approx \hat{I}(t)\hat{\phi}(t), \tag{47}$$

where the junction current operator is unchanged.

The interaction picture Hamiltonian of Eq. (47) is the starting point to derive a Markovian master equation for the resonator modes. Following the standard approach, this is done by going to second order in $\hat{H}_I$ and tracing out the junction degrees of freedom leading to [6]

$$\dot{\rho}(t) = -\frac{1}{\hbar^2} \int_0^\infty \mathrm{d}\tau \, \text{tr}_{\text{tj}}[\hat{H}_I(t), [\hat{H}_I(\tau), \rho(t) \otimes \rho_{\text{tj}}]]. \tag{48}$$

Here $\rho(t)$ is the density matrix of the resonator modes and $\rho_{\mathrm{tj}}$ of the tunnel junction. Eq. (48) can be expressed in terms of the one-sided Fourier transform

$$S(t,\omega) = \int_0^\infty \mathrm{d}\tau \langle \hat{I}(t)\hat{I}(t-\tau)\rangle \mathrm{e}^{i\omega\tau}, \tag{49}$$

where the brackets refer to an expectation value with respect to $\rho_{\mathrm{tj}}$. We find

$$
\begin{aligned}
\dot{\rho}(t) = \sum_{m,n} \Big\{ &-\frac{i}{\hbar}[\Delta(\omega_m,-\omega_n,t)\hat{a}_m^\dagger \hat{a}_n, \rho(t)] + \Gamma(\omega_m,-\omega_n,t)\mathcal{C}[\hat{a}_n,\hat{a}_m^\dagger]\rho(t) \\
&-\frac{i}{\hbar}[\Delta(-\omega_m,\omega_n,t)\hat{a}_m \hat{a}_n^\dagger, \rho(t)] + \Gamma(-\omega_m,\omega_n,t)\mathcal{C}[\hat{a}_n^\dagger,\hat{a}_m]\rho(t) \\
&-\frac{i}{\hbar}[\Delta(-\omega_m,-\omega_n,t)\hat{a}_m \hat{a}_n, \rho(t)] + \Gamma(-\omega_m,-\omega_n,t)\mathcal{C}[\hat{a}_n,\hat{a}_m]\rho(t) \\
&-\frac{i}{\hbar}[\Delta(\omega_m,\omega_n,t)\hat{a}_m^\dagger \hat{a}_n^\dagger, \rho(t)] + \Gamma(\omega_m,\omega_n,t)\mathcal{C}[\hat{a}_n^\dagger,\hat{a}_m^\dagger]\rho(t) \Big\},
\end{aligned}
\tag{50}
$$

where

$$\mathcal{C}[\hat{x},\hat{y}]\rho = \hat{x}\rho\hat{y} - \frac{1}{2}\{\hat{y}\hat{x},\rho\}, \tag{51}$$

and where we have defined the Lamb shifts and rates

$$\Delta(\omega_m,\omega_n,t) = \frac{1}{2i}\sqrt{Z_m Z_n}\, u_m(-L) u_n(-L) \mathrm{e}^{i(\omega_m+\omega_n)t} \left[ S(t,-\omega_n) - S(t,\omega_m)^* \right], \tag{52}$$

$$\Gamma(\omega_m,\omega_n,t) = \frac{1}{\hbar}\sqrt{Z_m Z_n}\, u_m(-L) u_n(-L) \mathrm{e}^{i(\omega_m+\omega_n)t} \left[ S(t,-\omega_n) + S(t,\omega_m)^* \right]. \tag{53}$$

Using the Fourier transformed current $\hat{I}[\omega] = \int_{-\infty}^\infty \mathrm{d}t \exp(i\omega t)\hat{I}(t)$, we can write

$$S(t,\omega) = \frac{1}{4\pi}\int_{-\infty}^\infty \mathrm{d}\omega' \langle \hat{I}[\omega']\hat{I}[-\omega]\rangle \mathrm{e}^{-i(\omega'-\omega)t}, \tag{54}$$

where we have used

$$\int_0^\infty \mathrm{d}\tau \mathrm{e}^{i(\omega''+\omega)\tau} = \pi\delta(\omega''+\omega) + iP\left(\frac{1}{\omega''+\omega}\right), \tag{55}$$

and we have dropped the principal part. Using Eqs. (13) and (54), the above rates $\Gamma$ and lamb shifts $\Delta$ can be taken to be time-independent by using the rotating-wave approximation and dropping all fast-rotating terms (assuming sufficiently high and well-separated mode frequencies). In the next three subsections, this will be done for particular choices of $\omega_{\mathrm{ac}}$.

### A. Single-mode squeezing

We first set $p\omega_{\mathrm{ac}} = 2\omega_m$. After dropping all rotating terms in Eq. (50), we are left with the following master equation for mode $m$

$$\dot{\rho}(t) = \kappa_m(N_m+1)\mathcal{D}[\hat{a}_m]\rho + \kappa_m N_m \mathcal{D}[\hat{a}_m^\dagger]\rho + \kappa_m M_m \mathcal{S}[\hat{a}_m]\rho + \kappa_m M_m^* \mathcal{S}[\hat{a}_m^\dagger]\rho, \tag{56}$$

where

$$\kappa_m(N_m+1) = \frac{Z_m u_m(-L)^2}{2\hbar} X_+^{(0)}(\omega_m), \tag{57}$$

$$\kappa_m N_m = \frac{Z_m u_m(-L)^2}{2\hbar} X_+^{(0)}(-\omega_m), \tag{58}$$

$$\kappa_m M_m = \frac{Z_m u_m(-L)^2}{2\hbar} X_+^{(p)}(\omega_m). \tag{59}$$

This can be expressed in terms of noise spectral densities by using

$$X_+^{(0)}(\omega_m) = \tilde{S}(\omega_m) + S_{\text{vac}}(\omega_m), \tag{60}$$

$$X_+^{(0)}(-\omega_m) = \tilde{S}(\omega_m) - S_{\text{vac}}(\omega_m), \tag{61}$$

$$X_+^{(p)}(\omega_m) = X(\omega_m) \quad (p > 0), \tag{62}$$

with $S_{\text{vac}}(\omega)$, $\tilde{S}(\omega)$ and $X(\omega)$ defined in Eqs. (18), (21) and (22), respectively. Note that Eq. (59) implies that $M_m$ is real. Using this, we obtain

$$\kappa_m = \frac{1}{\hbar} Z_m u_m(-L)^2 S_{\text{vac}}(\omega_m), \tag{63}$$

$$N_m = \frac{1}{2} \frac{\tilde{S}(\omega_m) - S_{\text{vac}}(\omega_m)}{S_{\text{vac}}(\omega_m)}, \tag{64}$$

$$M_m = \frac{1}{2} \frac{X(\omega_m)}{S_{\text{vac}}(\omega_m)}. \tag{65}$$

The above expression for the cavity damping rate $\kappa_m$ can be compared with the phenomenological model of Sec. III. Indeed, taking the large impedance mismatch limits of Eqs. (37) and (39) in Eq. (63), we find that $\kappa_m$ can be written as

$$\kappa = \frac{Z_0}{R} \frac{v}{L} \quad \text{for } \frac{Z_0}{R} \to 0, \tag{66}$$

$$\kappa = \frac{R}{Z_0} \frac{v}{L} \quad \text{for } \frac{Z_0}{R} \to \infty. \tag{67}$$

These expressions coincide with the imaginary parts of Eqs. (36) and (38) as expected.

From Eq. (56), the variance of the quadratures $\hat{X}_m = \hat{a}_m + \hat{a}_m^\dagger$ and $\hat{Y}_m = -i\hat{a}_m + i\hat{a}_m^\dagger$ in steady-state are

$$\Delta X_m^2 = 2\left(N_m + \frac{1}{2} - M_m\right) = \frac{\tilde{S}(\omega_m) - X(\omega_m)}{S_{\text{vac}}(\omega_m)}, \tag{68}$$

$$\Delta Y_m^2 = 2\left(N_m + \frac{1}{2} + M_m\right) = \frac{\tilde{S}(\omega_m) + X(\omega_m)}{S_{\text{vac}}(\omega_m)}, \tag{69}$$

where $\Delta O^2 = \langle \hat{O}^2 \rangle - \langle \hat{O} \rangle^2$. The quadrature $\hat{X}_m$ is thus squeezed if $\tilde{S}(\omega_m) - X(\omega_m) < S_{\text{vac}}(\omega_m)$. Note, however, that the Heisenberg uncertainty principle implies that $\Delta X_m^2 \Delta Y_m^2 \geq 1$, or in other words

$$\tilde{S}(\omega_m) - X(\omega_m) \geq \frac{S_{\text{vac}}(\omega_m)^2}{\tilde{S}(\omega_m) + X(\omega_m)}, \tag{70}$$

which puts a bound on the degree of squeezing.

One can also consider the output field from the resonator when adding another decay channel, $e.g.$ by considering a capacitive coupling to an output at $x = 0$ as illustrated in Fig. 1 of the Letter. Weak coupling to this output port only slightly modifies the mode envelopes and adds new dissipative terms to the master equation. For simplicity, taking the field in this new output channel to be in the vacuum state, and calling the damping rate associated to this channel $\kappa'$, we find that the squeezing of the $\hat{X}_m$ quadrature is modified to be

$$\Delta X_m^2 = \frac{2\kappa_m}{\kappa_m + \kappa_m'}(N_m + \frac{1}{2} - M_m) + \frac{\kappa_m'}{\kappa_m + \kappa_m'}. \tag{71}$$

As usual, the last term shows that the squeezing is degraded due to the vacuum noise from the new output port. It is also interesting to compare the squeezing of the intracavity field to that of the output field emitted through the output port. To do this it is convenient to compute the squeezing spectrum of the intracavity and output fields, respectively. The squeezing spectra are defined as

$$S_{\text{intra}}[\omega] = \frac{1}{2\pi} \int_{-\infty}^{\infty} d\omega' \langle \hat{X}[\omega]\hat{X}[\omega'] \rangle, \tag{72}$$

$$S_{\text{out}}[\omega] = \frac{1}{2\pi} \int_{-\infty}^{\infty} d\omega' \langle \hat{X}_{\text{out}}[\omega]\hat{X}_{\text{out}}[\omega'] \rangle, \tag{73}$$

where $\hat{X}_{\text{out}}[\omega]$ is the Fourier transformed output field [6]. We find using standard methods [6] that

$$S_{\text{intra}}[\omega] = \frac{2\kappa_m}{\omega^2 + (\kappa'_m + \kappa_m)^2/4}(N_m + \tfrac{1}{2} - M_m) + \frac{\kappa'_m}{\omega^2 + (\kappa'_m + \kappa_m)^2/4}, \tag{74}$$

$$S_{\text{out}}[\omega] = \frac{2\kappa'_m \kappa_m}{\omega^2 + (\kappa'_m + \kappa_m)^2/4}(N_m + \tfrac{1}{2} - M_m) + \frac{\omega^2 + (\kappa'_m - \kappa_m)^2/4}{\omega^2 + (\kappa'_m + \kappa_m)^2/4}. \tag{75}$$

An interesting case is when $\kappa'_m = \kappa_m$ where the squeezing spectra at $\omega = 0$ are $S_{\text{intra}}[\omega = 0] = 2(N_m + \tfrac{1}{2} - M_m + \tfrac{1}{2})$ and $S_{\text{out}}[\omega = 0] = 2(N_m + \tfrac{1}{2} - M_m)$. This shows that the squeezing of the intracavity field is degraded by the vacuum noise of the new decay channel ($\Delta X_m^2$ cannot go below $1/2$), while this is not the case for the output field. The squeezing spectrum of the output field at $\omega = 0$ for $\kappa'_m = \kappa_m$ is equal to that of the intracavity field in the limit $\kappa'_m \to 0$ (with $\kappa_m > 0$).

## B. Two-mode squeezing

We now consider a setup where the tunnel junction is ac biased such that $p\omega_{\text{ac}} = \omega_m + \omega_n$ for an integer $p$ and two modes $m$ and $n$. We find after dropping all rotating terms

$$\begin{aligned}
\dot{\rho}(t) = \sum_{l=m,n} &\left\{ \kappa_l(N_l + 1)\mathcal{D}[\hat{a}_l]\rho + \kappa_l N_l \mathcal{D}[\hat{a}_l^\dagger]\rho \right\} \\
&+ \sqrt{\kappa_m \kappa_n} M_{mn}^* \mathcal{C}[\hat{a}_n, \hat{a}_m]\rho + \sqrt{\kappa_m \kappa_n} M_{mn} \mathcal{C}[\hat{a}_m^\dagger, \hat{a}_n^\dagger]\rho \\
&+ \sqrt{\kappa_m \kappa_n} M_{nm}^* \mathcal{C}[\hat{a}_m, \hat{a}_n]\rho + \sqrt{\kappa_m \kappa_n} M_{nm} \mathcal{C}[\hat{a}_n^\dagger, \hat{a}_m^\dagger]\rho,
\end{aligned} \tag{76}$$

where $\kappa_l$ and $N_l$ are defined as before in Eqs. (63)–(64), for $l = m, n$, while $M_{m,n}$ is

$$\sqrt{\kappa_m \kappa_n} M_{mn} = \frac{\sqrt{Z_m Z_n}\, u_m(-L) u_n(-L)}{4\pi\hbar} X_+^{(p)}(\omega_n) \tag{77}$$

$$\sqrt{\kappa_m \kappa_n} M_{nm} = \frac{\sqrt{Z_m Z_n}\, u_m(-L) u_n(-L)}{4\pi\hbar} X_+^{(p)}(\omega_m). \tag{78}$$

Using again that $X_+^{(p)}(\omega)$ is real, and the fact that $\langle \hat{I}[\omega_n]\tilde{I}[\omega_m] \rangle = \langle \hat{I}[\omega_m]\tilde{I}[\omega_n] \rangle$, which implies that $M_{mn} = M_{nm}$ (real) [2], we can then write for the two-mode master equation

$$\begin{aligned}
\dot{\rho}(t) = \sum_{l=n,m} &\left\{ \kappa_l(N_l + 1)\mathcal{D}[\hat{a}_l]\rho + \kappa_l N_l \mathcal{D}[\hat{a}_l^\dagger]\rho \right\} \\
&+ \sqrt{\kappa_n \kappa_m} M_{nm} \left( \hat{a}_n \rho \hat{a}_m + \hat{a}_m \rho \hat{a}_n + \{\hat{a}_n \hat{a}_m, \rho\} \right) \\
&+ \sqrt{\kappa_n \kappa_m} M_{nm} \left( \hat{a}_n^\dagger \rho \hat{a}_m^\dagger + \hat{a}_m^\dagger \rho \hat{a}_n^\dagger + \{\hat{a}_n^\dagger \hat{a}_m^\dagger, \rho\} \right),
\end{aligned} \tag{79}$$

where

$$M_{nm} = \frac{1}{2} \frac{X(\omega_n)}{\sqrt{S_{\text{vac}}(\omega_n) S_{\text{vac}}(\omega_m)}}. \tag{80}$$

To consider two-mode squeezing we define the following two-mode quadratures

$$\hat{X}_\pm = \hat{X}_n \pm \hat{X}_m, \tag{81}$$

$$\hat{Y}_\pm = \hat{Y}_n \pm \hat{Y}_m. \tag{82}$$

Assuming for simplicity that $\kappa_n = \kappa_m$, we find that the variances in steady state are

$$\Delta X_+^2 = \Delta Y_-^2 = 2(N_n + N_m + 1 - 2M_{nm}), \tag{83}$$

$$\Delta X_-^2 = \Delta Y_+^2 = 2(N_n + N_m + 1 + 2M_{nm}). \tag{84}$$

The commuting two-mode quadratures $\hat{X}_+$ and $\hat{Y}_-$ are both squeezed for $2M_{nm} > N_n + N_m$.

## C. Two-mode photon conversion

Finally, it is interesting to consider the influence of the noise emitted from the junction when the ac bias frequency is matched to the frequency difference of two modes, *i.e.*, $p\omega_{ac} = \omega_n - \omega_m$, for $p$ a positive integer ($\omega_n > \omega_m$). Once more dropping all rotating terms we find

$$
\begin{aligned}
\dot{\rho}(t) = \sum_{l=n,m} & \left\{ \kappa_l(N_l+1)\mathcal{D}[\hat{a}_l]\rho + \kappa_l N_l \mathcal{D}[\hat{a}_l^\dagger]\rho \right\} \\
& + \sqrt{\kappa_m \kappa_n} T_{mn} \mathcal{C}[\hat{a}_n, \hat{a}_m^\dagger]\rho + \sqrt{\kappa_m \kappa_n} T_{mn}^* \mathcal{C}[\hat{a}_m, \hat{a}_n^\dagger]\rho \\
& + \sqrt{\kappa_m \kappa_n} T_{nm} \mathcal{C}[\hat{a}_m, \hat{a}_n^\dagger]\rho + \sqrt{\kappa_m \kappa_n} T_{nm}^* \mathcal{C}[\hat{a}_n, \hat{a}_m^\dagger]\rho.
\end{aligned}
\tag{85}
$$

This master equation describes a dissipative process where photons can be converted from mode $n$ to $m$ and vice versa. The parameter $T_{mn}$ is given by

$$
T_{mn} = \frac{\sqrt{Z_m Z_n} u_m(-L) u_n(-L)}{4\pi\hbar} X_+^{(p)}(\omega_n),
\tag{86}
$$

$$
T_{nm} = \frac{\sqrt{Z_m Z_n} u_m(-L) u_n(-L)}{4\pi\hbar} X_+^{(p)}(\omega_n),
\tag{87}
$$

Again, we have that $T_{mn} = T_{nm} \equiv M_{nm}$, where $M_{nm}$ is given by Eq. (80). This leads to

$$
\begin{aligned}
\dot{\rho}(t) = \sum_{l=n,m} & \left\{ \kappa_l(N_l+1)\mathcal{D}[\hat{a}_l]\rho + \kappa_l N_l \mathcal{D}[\hat{a}_l^\dagger]\rho \right\} \\
& + \sqrt{\kappa_n \kappa_m} M_{nm} \left( \hat{a}_n \rho \hat{a}_m^\dagger + \hat{a}_m^\dagger \rho \hat{a}_n + \{\hat{a}_n \hat{a}_m^\dagger, \rho\} \right) \\
& + \sqrt{\kappa_n \kappa_m} M_{nm} \left( \hat{a}_n^\dagger \rho \hat{a}_m + \hat{a}_m \rho \hat{a}_n^\dagger + \{\hat{a}_n^\dagger \hat{a}_m, \rho\} \right).
\end{aligned}
\tag{88}
$$

In steady-state under the above master equation, and for $\kappa_m = \kappa_n$ for simplicity, the variances of the two-mode quadratures defined above are now

$$
\Delta X_-^2 = \Delta Y_-^2 = 2(N_n + N_m + 1 - 2M_{nm}),
\tag{89}
$$

$$
\Delta X_+^2 = \Delta Y_+^2 = 2(N_n + N_m + 1 + 2M_{nm}).
\tag{90}
$$

Importantly, the two "squashed" quadratures, $\hat{X}_-$ and $\hat{Y}_-$ are non-commuting, with $[\hat{X}_-, \hat{Y}_-] = 4i$, implying the following uncertainty relation

$$
\Delta X_- \Delta Y_- \geq \frac{1}{2} |[U_-, V_-]| = 2,
\tag{91}
$$

which leads to

$$
2M_{nm} \leq N_a + N_b.
\tag{92}
$$

In other words, the quadrature variances cannot be squeezed below their vacuum value. The two modes become correlated through the bath-induced photon conversion process ($\langle \hat{a}_n^\dagger \hat{a}_m \rangle$ is non-zero in steady state), but they do not become entangled through this process.

---



\end{document}